\documentclass[twoside,twocolumn,english,pra,twocolumn,aps,superscriptaddress,showpacs]{revtex4}
\usepackage{slashbox}
\usepackage[T1]{fontenc}
\usepackage[latin9]{inputenc}
\usepackage{color}
\usepackage{babel}
\usepackage{textcomp}
\usepackage{amsmath}
\usepackage{amsthm}
\usepackage{graphicx}
\usepackage[unicode=true,pdfusetitle,
 bookmarks=true,bookmarksnumbered=false,bookmarksopen=false,
 breaklinks=false,pdfborder={0 0 0},backref=false,colorlinks=true]
 {hyperref}
\usepackage{breakurl}
\usepackage{epstopdf}
\usepackage{amsmath,bm}

\newcommand{\ehbar}{\hbar_{\mathrm{eff}}}

\begin{document}

\title{Coexistence of directed momentum current and ballistic energy diffusion in coupled non-Hermitian kicked rotors}

\author{Jian-Zheng Li}
\affiliation{School of Science, Jiangxi University of Science and Technology, Ganzhou 341000, China}
\author{Wen-Lei Zhao}
\email{wlzhao@jxust.edu.cn}
\affiliation{School of Science, Jiangxi University of Science and Technology, Ganzhou 341000, China}
\author{Jie Liu}
\email{jliu@gscaep.ac.cn}
\affiliation{Graduate School, China Academy of Engineering Physics, Beijing 100193, China}
\affiliation{HEDPS, Center for Applied Physics and Technology, and College of Engineering, Peking University, Beijing 100871, China}

\begin{abstract}
We numerically investigate the quantum transport in a coupled kicked rotors with the $\mathcal{PT}$-symmetric  potential. We find that the spontaneous $\mathcal{PT}$-symmetry breaking of wavefunctions emerges when the amplitude of the imaginary part of the complex potential is beyond a threshold value, which can be  modulated by the coupling strength effectively. In the regime of the $\mathcal{PT}$-symmetry breaking, the particles driven by the periodical kicks move unidirectionally in momentum space, indicating the emergence of a directed current. Meanwhile, with increasing the coupling strength, we find a transition from the ballistic energy diffusion to a kind of the modified ballistic energy diffusion where the width of the wavepacket also increases with time in a power law. Our findings suggest that the decoherence effect induced by the interplay between the inter-particle coupling and the non-Hermitian driving potential is responsible for these particular transport behaviors.
\end{abstract}
\date{\today}
\maketitle

\section{Introduction}\label{intro}
Directed transport and quantum diffusion in both real and momentum space have attracted many interests in diverse fields of physics, ranging from condensed matter physics~\cite{Cooper19} to quantum chaos~\cite{JWang20,LWang07}, and to biological physics~\cite{Fornes21}. It is found that a seminal phenomenon of directed transport, namely, quantum ratchet~\cite{Arzola17}, has practical applications in the design of electron pumps~\cite{Lau20}, in enhancing the efficiency of photovoltaic cells~\cite{Sogabe21} and in the construction of molecular rotors~\cite{Denisov14}.
In other aspects, the quantum diffusion is revelent for understanding the conductivity of electronics~\cite{Anderson58}, the spin transport~\cite{Nardis21,Ljubotina19}, the energy transport~\cite{Uchiyama18}, as well as the information scrambling~\cite{Moudgalya19,Lewis19}, thus has been a subject of intense study in various areas of physics.
A landmark of the study on quantum diffusion is the Anderson localization (AL) of electrons in disordered potential~\cite{Anderson58,Lagendijk09}. Its analog in momentum space is the dynamical localization (DL) in the quantum kicked rotor (QKR) which is a paradigmatic model of Floquet systems~\cite{Casati79,Fishman82}. The QKR model with incommensurable frequency can mimic the Anderson model in 2D or 3D disordered lattice, and is very convenient for the theoretical investigation and experimental realization~\cite{Shepelyansky83,Casati89,Lemarie08,Lemarie09}.

Nowadays, the QKR and its variants have been accepted as ideal systems for exploring rich physics, for instance, Floquet-topological phase~\cite{Shirley65,Bukova15,Ho12,YuChen14,Longwen18}, dynamical phase transition~\cite{Hainaut18}, and quantum walk in momentum-space lattice~\cite{Dadras18,Dadras19,Xie20}.
More recently, the study of the $\cal{PT}$-symmetric extension of the kicked rotor (PTKR) shows the spontaneous $\cal{PT}$-symmetry breaking characterized by the emergence of the complex quasienergies of the Floquet operators~\cite{Longhi17,West10}. Interestingly, the $\mathcal{PT}$-symmetric kicking potential leads to quantized acceleration of momentum current~\cite{Zhao19,Zhao20cpb} and quantized response of out-of-time ordered correlators~\cite{Zhaowl22} in QKR model. The non-Hermiticity of Hamiltonian has nowadays been widely accepted as a fundamental modification for the conventional quantum mechanics~\cite{Bender98,Ashida20,Moiseyev11}. It is known that open systems which exchange particles or energy with environment can be described by non-Hermitian Hamiltonians~\cite{Ritsch13,Makris08,Li20}, for instance ultracold atoms in dissipative optical lattice, optical wave propagation in lossy media, electrical circuits with virtual absorption, just name a few.

\begin{figure}[htbp]
\begin{center}
\includegraphics[width=8.0cm]{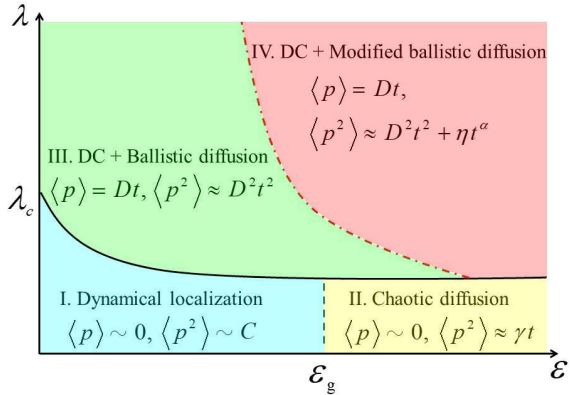}
\caption{Schematic diagram for the directed current (DC) and quantum diffusion in the parameter space $(\lambda,\varepsilon)$. \label{PDiagram}}
\end{center}
\end{figure}

The quantum transport in non-Hermitian systems has received intensive investigations~\cite{Eichelkraut13,Xu21}, where the fate of the directed current (DC) and DL under the effects of inter-particle coupling is still an open issue~\cite{Borgonovi16,Lellouch20,Chicireanu21,Vuatelet21,Casati06,Duval,Park03}. It is found that in a system of coupled QKRs, the non-Hermitian driving potential can protect the DL, which otherwise is destroyed by the inter-particle coupling in Hermitian case~\cite{Huang21}. In this paper, we investigate the effects of interaction on the quantum transport in momentum space via a system of coupled PTKRs. Interestingly, we find the emergence of the spontaneous $\mathcal{PT}$-symmetry breaking when the strength of the imaginary parts of the complex kicking potential is beyond a threshold value, which can be effectively modulated by the coupling strength. In the regime where the $\mathcal{PT}$-symmetry is unbroken, there are two different phases of energy diffusion, one is the DL, i.e., $\langle p_1^2 \rangle\sim C$ and another is the chaotic diffusion $\langle p_1^2\rangle \approx \gamma t$, both of which have no the phenomenon of the DC, i.e., $\langle p_1 \rangle\sim 0$ (see phases I and II in Fig.~\ref{PDiagram}). In the regime of $\mathcal{PT}$-symmetry breaking, each KR moves unidirectionally in momentum space, i.e., $\langle p_1 \rangle=Dt$, indicating the appearance of the DC. Interestingly, there also two different phases of energy diffusion, one is the ballistic diffusion, i.e., $\langle p_1^2 \rangle\approx D^2t^2$ (see phase III in Fig.~\ref{PDiagram}), and another is the modified ballistic diffusion (MBD), i.e., $\langle p_1^2 \rangle\approx D^2t^2 + \eta t^{\alpha}$ (see phase IV in Fig.~\ref{PDiagram}).
The coexistence of the DC and the MBD is a unique phenomenon in coupled non-Hermitian systems. We numerically obtain the acceleration rate $D$ of DC and the diffusion rate $\eta$ of energy for a wide regime of system parameters, which is helpful to guide the Floquet engineering of transport in momentum-space lattice~\cite{Keser16,Hainaut19}. We also numerically find that entanglement grows with the increase of the strength of interaction. It is reasonable to believe that the coexistence of the DC and MBD results from the intrinsic decoherece effects in non-Hermitian chaotic systems.

The paper is organized as follows. In Sec.~\ref{SEC-MR}, we describe the system. In Sec.~\ref{Tranbehavior}, we show the transport behaviors in our system with emphasis on the coexistence of the DL and MBD. A summary is presented in Sec.~\ref{SEC-SUM}.

\section{Model}\label{SEC-MR}
The Hamiltonian of the coupled PTKRs reads
\begin{equation}\label{Hamil}
  \text{H}=\text{H}_{1}+\text{H}_{2}+\text{H}_{\rm{I}}\;,
\end{equation}
with $\text{H}_{j}$ ($j=1,2$) of the individual particle
\begin{equation}
  \text{H}_{j}=\frac{p_{j}^2}{2}+V(\theta_{j})\sum^\infty_{n=0}\delta(t-t_{n})\;,
\end{equation}
and the inter-particle coupling
\begin{equation}
  \text{H}_{\text{I}}=\varepsilon\cos(\theta_{1}-\theta_{2})\sum^\infty_{n=0}\delta(t-t_{n}).
\end{equation}
Here, our consideration of the temporally delta modulation of the coupling is just for the convenience of numerical simulation. This kind of coupling has been widely used in previous investigations~\cite{Rozenbaum17,Duval}. In our system, the kicking potential is $\cal{PT}$-symmetric, i.e., $V(\theta_{j})=K[\cos(\theta_{j})+i\lambda \sin(\theta_{j})]$ with $K$ being the kick strength while $\lambda$ controlling the strength of imaginary part of $V(\theta_{j})$. The $p_{j}=-i\hbar_{\rm{eff}}\partial/\partial\theta_{j}$ is the angular momentum operator and $\theta_{j}$ is the angle coordinate of each subsystem. The $\hbar_{\rm{eff}}$ indicates the effective Planck constant, and the parameter $\varepsilon$ is the coupling strength. The time $t_{n}(=0, 1,\ldots)$ is integer indicating the number of kicks. All variables are properly scaled and thus in dimensionless units.

The eigenequation of $p_j$ has the expression $p_{j}|\phi^j_{m}\rangle=p^j_{m}|\phi^j_{m}\rangle$ with $p^j_{m}=m\hbar_{\rm{eff}}$ and $\langle \theta_j|\phi^j_{m}\rangle=e^{im\theta_j}/\sqrt{2\pi}$.
On the basis of the product states
$|\phi^1_{m},\phi^2_{n}\rangle=|\phi^1_{m}\rangle\otimes|\phi^2_{n}\rangle$, an arbitrary quantum state $|\psi\rangle$ can be expanded as $|\psi\rangle=\sum_{m,n}\psi_{m,n}|\phi^1_{m},\phi^2_{n}\rangle$. One period evolution of the quantum state from $t_{n}$ to $t_{n+1}$ is given by $|\psi(t_{n+1})\rangle=U|\psi(t_{n})\rangle$. The Floquet operator $U$ can be separated into two fractions,
\begin{equation}
  U=U_{\rm_{f}}U_{\rm_{K}}\;,
\end{equation}
where the free evolution operator of the kinetic term takes the form
\begin{equation}
  U_{\rm_{f}}=\exp\left(-\frac{i}{\hbar_{\rm_{eff}}}\sum^2_{j=1}\frac{p_{j}^2}{2}\right)\;,
\end{equation}
and the evolution operator of the kicking term is
\begin{equation}
  U_{\rm_{K}}=\exp \left \{-\frac{i}{\hbar_{\rm_{eff}}}\left[\sum^2_{j=1}V(\theta_{j})+\varepsilon\cos(\theta_{1}-\theta_{2})\right]\right \}\;.
\end{equation}
In our investigations, the initial state is set to be the product of the ground states, i.e., $|\psi(t_{0})\rangle=|\phi^1_{0},\phi^2_{0}\rangle$.

\section{Directed current and energy diffusion in the momentum space}\label{Tranbehavior}

\subsection{Time evolution of $\langle p_1\rangle$, $\langle p_1^2\rangle$ and $\mathcal{M}_1$}

The transport behavior of one of particles (say particle 1) in momentum space is characterized by the momentum current $\langle p_{1}\rangle=\rm{Tr}(\rho_{1}\emph{p}_{1})$, the energy diffusion $\langle p^{2}_{1}\rangle=\rm{Tr}(\rho_{1}\emph{p}^2_{1})$, and the width of the time-evolved state $\mathcal{M}_1=\langle p^{2}_{1}\rangle-(\langle p_{1}\rangle)^{2}$. In the emergence of $\cal{PT}$-symmetry breaking, the norm $\cal{N}$ of quantum state exponentially increases with time. To eliminate the contribution of the norm to the observable, we normalized the quantum state to be unity after each kick. Therefore, we define the reduced density matrix of particle 1 as $\rho_{1}=\frac{1}{\mathcal{N}}\rm{Tr}_{2}(|\psi\rangle\langle\psi|)$, which is obtained by tracing out the other degree of freedom from the density
matrix of the two-particle system $\rho=|\psi\rangle\langle\psi|$.

We show the phase diagram of transport behaviors for different $\lambda$ and $\varepsilon$ in Fig.~\ref{PDiagram}. In the unbroken phase of $\cal{PT}$-symmetry, i.e., $\lambda<\lambda_c$, there are two different classes of transport behaviors, both of which have no the DC, i.e., $\langle p_{1}\rangle\sim0$. Class I corresponds to DL, i.e., $\langle p_{1}^2\rangle\sim C$ with $\varepsilon<\varepsilon_g$, and class II is for the chaotic diffusion, i.e., $\langle p_{1}^2\rangle\approx \gamma t$ with $\varepsilon>\varepsilon_g$. In the regime of $\cal{PT}$-symmetry breaking, i.e., $\lambda>\lambda_c$, there are also two different classes. In class III, the system displays the DC $\langle p_{1}\rangle = D t$~\cite{Zhao19} and ballistic diffusion $\langle p^2_{1}\rangle \approx D^2 t^2$~\cite{Zhaowl22}. The corresponding width of quantum state is almost a constant with time evolution, as the wavepacket is a soliton moving unidirectionally in momentum space~\cite{Zhao20cpb}. Our main finding in this work is the exotic phenomenon of the coexistence of the DC and MBD in class IV. Specifically, the momentum current linearly increases with time
\begin{equation}\label{LinCurret}
  \langle p_{1}\rangle=Dt\;.
\end{equation}
The energy diffuses in a MBD of time
\begin{equation}\label{Engdiffu}
  \langle p^{2}_{1}\rangle\approx D^2t^{2}+\eta t^{\alpha}\quad \text{with} \quad \alpha< 2\;.
\end{equation}
Correspondingly, the width of quantum state grows in the power-law of time
\begin{equation}\label{VARiance}
  \mathcal{M}_1=\eta t^{\alpha}\;.
\end{equation}
Here, the coefficients $D$ and $\eta$, as well as the exponent $\alpha$ depend on the system parameters.

\begin{figure}[htbp]
\begin{center}
\includegraphics[width=8.5cm]{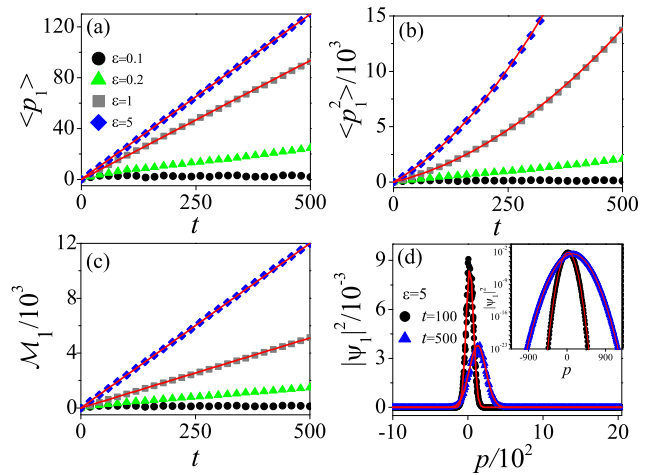}
\caption{Time evolution of $\langle p_{1}\rangle$ (a), $\langle p^{2}_{1}\rangle$ (b), and $\mathcal{M}_1$ (c) with $\varepsilon=0$ (circles), 0.2 (triangles), 1 (squares), and 5 (diamonds). Red lines in (a), (b) and (c) indicate $\langle p_{1}\rangle=Dt$ in Eq.~\eqref{LinCurret}, $\langle p^{2}_{1}\rangle\approx D^{2}t^{2}+\eta t^{\alpha}$ in Eq.~\eqref{Engdiffu}, and $\mathcal{M}_1=\eta t^{\alpha}$ in Eq.~\eqref{VARiance} with $\alpha=1$. (d) Momentum distributions $|\psi_{1}(p)|^{2}$ at the time $t=100$ (circles) and $500$ (triangles) for $\varepsilon=5$. Red lines indicate the fitting functions of the Gaussian form $|\psi_{1}(p)|^{2}\propto e^{-(p-p_{c})^{2}/\sigma}$. Inset: same as in the main plot but on a logarithmic y scale. The parameters are $K=5$, $\ehbar=1$, and $\lambda=0.01$.\label{Diffusion}}
\end{center}
\end{figure}

In the regime of the unbroken phase of $\cal{PT}$-symmetry with real spectrum of quasienergies, the transport behavior of this system has no essential differences with that of Hermitian case. In our numerical investigations, we choose very small $\lambda$, so that the $\cal{PT}$-symmetry breaking does not occur with $\varepsilon=0$. Indeed, our numerical results with $\varepsilon=0$ show that there is neither momentum current, i.e., $\langle p_1\rangle\sim0$ [see Fig.~\ref{Diffusion}(a)], nor energy diffusion, i.e., $\langle p^2_1\rangle\sim C$ [see Fig.~\ref{Diffusion}(b)], correspondingly $\mathcal{M}_1\sim C$ [see Fig.~\ref{Diffusion}(c)], which is a clear evidence of the appearance of DL.
In the regime of the $\cal{PT}$-symmetry breaking, the system exhibits exotic transport behaviors. We find that the mean value $\langle p_1\rangle$ increases linearly with time, i.e., $\langle p_1\rangle=Dt$ [e.g., $\varepsilon=1$ in Fig.~\ref{Diffusion}(a)], which demonstrates the emergence of DC. Meanwhile, the energy diffusion increases in a way of MBD $\langle p^2_1\rangle \approx D^2 t^2 +\eta t^{\alpha} $ with $\alpha=1$ [e.g., $\varepsilon=1$ in Fig.~\ref{Diffusion}(b)]. The corresponding width of the time-evolved wavepacket grows as $\mathcal{M}_1=\eta t^{\alpha}$ [e.g., $\varepsilon=1$ in Fig.~\ref{Diffusion}(c)], indicating the fact that the unbounded spreading of wavepacket occurs. Our results, therefore, present a solid evidence of the coexistence of the DC and MBD due to the interplay between the non-Hermitian driving and the coupling.
After extensive investigations on the energy diffusion for different $\lambda$, we find that $\alpha$ varies with $\lambda$, which demonstrates the influences of the non-Hermitian driven potential on the energy diffusion.

The probability density distributions of particle 1 in momentum space are shown in Fig.~\ref{Diffusion}(d). One can see that the momentum distribution can be well described by the Gaussian function, i.e., $|\psi_{1}(p,t)|^{2}\propto e^{-[p-p_{c}(t)]^{2}/\sigma(t)}$. Interestingly, the center $p_{c}(t)$ of Gaussian wavepacket increases with time, which reveals the emergence of the DC in momentum space. Moreover, its width $\sigma(t)$ also increases with time, corresponding to the unbound growth of $\mathcal{M}_1(t)$. The appearance of Gaussian distribution is usually regarded as a signature of the loss of quantum coherence~\cite{Gadway13,Casati06} which results in the exponentially-localized quantum states, namely, a character of DL~\cite{Fishman82,Casati89,Casati79}, in momentum space.
Previous investigations on Hermitian systems have reported that the coupling induces the spreading of the Gaussian wavepacket with time, while its center $p_c$ is fixed, thus no DC.
Our finding of the coexistence of the increase of both $\sigma$ and $p_c$ is a new kind transport phenomenon due to the quantum decoherence effects in non-Hermitian chaotic systems.
\begin{figure}[htbp]
\begin{center}
\includegraphics[width=8.5cm]{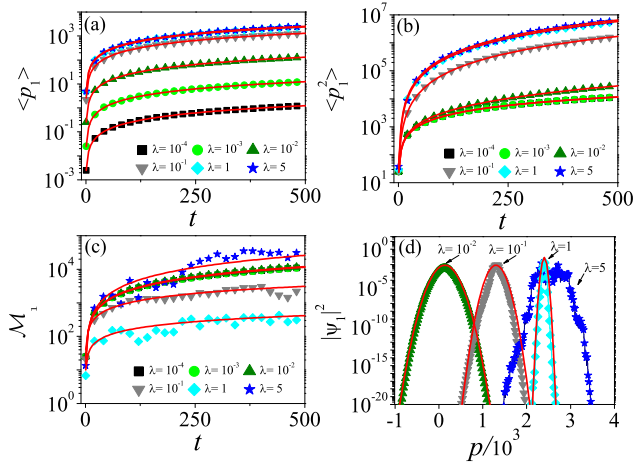}
\caption{Time dependence of $\langle p_{1}\rangle$ (a), $\langle p^{2}_{1}\rangle$ (b), and $\mathcal{M}_1$ (c) for $\varepsilon=5$ with $\lambda=10^{-4}$ (squares), $10^{-3}$ (circles), $0.01$ (up triangles), $0.1$ (down triangles), $1$ (diamonds), and $5$ (pentagrams). Red lines in (a), (b) and (c) indicate $\langle p_{1}\rangle=Dt$ in Eq.~\eqref{LinCurret}, $\langle p^{2}_{1}\rangle\approx D^{2}t^{2}+\eta t^{\alpha}$ in Eq.~\eqref{Engdiffu}, and $\mathcal{M}_1=\eta t^{\alpha}$ in Eq.~\eqref{VARiance} with $\alpha<2$. (d) Momentum distributions $|\psi_{1}(p)|^{2}$ for $t=500$ and $\varepsilon=5$ with $\lambda=0.01$ (up triangles), $\lambda=0.1$ (down triangles), $\lambda=1$ (diamonds), and $\lambda=5$ (pentagrams). Red lines indicate the fitting function of the Gaussian form $|\psi_1(p)|^2\propto e^{-(p-p_c)^2/\sigma}$. Other parameters are same as in Fig.~\ref{Diffusion}.\label{Diffusion2}}
\end{center}
\end{figure}

We further numerically investigate the directed transport and energy diffusion for different $\lambda$ when the coupling $\varepsilon$ is sufficiently strong so that the $\cal{PT}$-symmetry phase breaking easily emerges for very small $\lambda$. Figure~\ref{Diffusion2}(a) shows that the momentum current linearly increases with time, i.e., $\langle p_{1}\rangle=Dt$. Meanwhile, the energy diffuses in a kidn of MBD $\langle p^{2}_{1}\rangle\approx D^{2}t^{2}+\eta t^{\alpha}$, for which both $\eta$ and $\alpha$ vary with $\lambda$ [see Fig.~\ref{Diffusion2}(b)]. Correspondingly, the width of wavepacket increases unboundedly $\mathcal{M}_1 = \eta t^{\alpha}$ [see Fig.~\ref{Diffusion2}(c)]. The momentum distributions are shown in Fig.~\ref{Diffusion2}(d). One can see that for weak non-Hermitian driving [e.g., $\lambda\leq1$ in Fig.~\ref{Diffusion2}(d)] the momentum distribution can be well described by the Gaussian function. However, for sufficiently large $\lambda$ [e.g., $\lambda=5$ in Fig.~\ref{Diffusion2}(d)], the quantum state is clearly different from the Gaussian wavepacket with irregular distribution in momentum space. It is reasonable to believe that the interplay between non-Hermiticity and coupling dramatically affects the decoherence effects, which leads to the irregular form of the momentum distribution.

\subsection{Growth rate of $\langle p_1\rangle$ and $\mathcal{M}_1$}
\begin{figure}[htbp]
\begin{center}
\includegraphics[width=7.5cm]{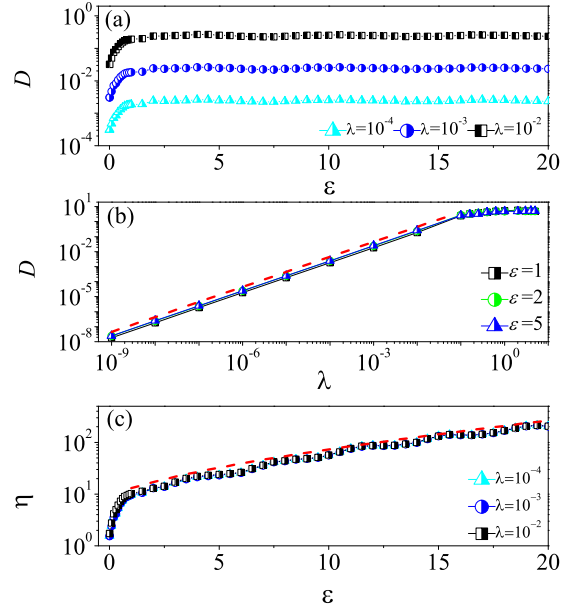}
\caption{(a) Growth rate $D$ versus $\varepsilon$ with $\lambda=10^{-4}$ (triangles), $\lambda=10^{-3}$ (circles), and $\lambda=10^{-2}$ (squares). (b) The value of $D$ versus $\lambda$ for $\varepsilon=1$ (squares), 2 (circles), and 5 (triangles). Red dashed line indicates the fitting function of the form $D\propto \lambda$. (c) Dependence of the $\eta$ on $\varepsilon$ with $\lambda=10^{-4}$ (triangles), $\lambda=10^{-3}$ (circles), and $\lambda=10^{-2}$ (squares). Red dashed line indicates the exponential fitting, i.e., $\eta\propto e^{\beta\varepsilon}$ with $\beta=0.1$. Other parameters are the same as in Fig.~\ref{Diffusion}.\label{Coefficient}}
\end{center}
\end{figure}

The growth rates of the momentum current and the width of quantum state are separately defined by $D=\langle p_1(t_f)\rangle/t_f$ and $\eta= \mathcal{M}_1(t_f)/t_f^{\alpha}$. In numerical simulations, the $t_f$ on the scale of hundreds of
kicking periods is enough to assure the high precision of numerical results. Figure~\ref{Coefficient}(a) shows that the $D$ increases rapidly from a very small value to saturation with increasing $\varepsilon$. Note that the nonzero value of $D$ for $\varepsilon=0$ is due to the finite time $t_f$ in numerical calculations. The saturation value of $D$ increases with the increase of $\lambda$, which reveals that the acceleration of momentum current is only determined by the non-Hermitian driving with no relation to coupling. As a further step, we numerically investigate the acceleration rate $D$ for various $\lambda$. Figure~\ref{Coefficient}(b) shows that the value of $D$ increases linearly with $\lambda$, i.e., $D\propto \lambda$, up to the saturation. Moreover, the $D$ is almost not dependent on the variation of $\varepsilon$ if the coupling strength is strong enough. The growth rate $\eta$ of the $\mathcal{M}_1$ for a wide regime of $\varepsilon$ and $\lambda$ is shown in Fig.~\ref{Coefficient}(c). One can find that the $\eta$ exponentially increases with $\varepsilon$, but without dependence on the variation of $\lambda$. Therefore, the spreading of wavepacket in momentum space is mainly determined by the inter-particle coupling. Since the MBD $\langle p^2_1\rangle \approx D^2t^2 + \eta t^{\alpha}$ has two parts, it is clear that the first part of quadratic growth origins from the non-Hermitian driving, while the second part, i.e., $\eta t^{\alpha}$ is dominated by the competition between coupling and non-Hermitian driving potential. Accordingly, this opens an opportunity for the experimental engineering of the transport behaviors in momentum-space lattice~\cite{Dadras18,Dadras19,Xie20}.

\subsection{Spontaneous $\cal{PT}$-symmetry breaking}
\begin{figure}[htbp]
\begin{center}
\includegraphics[width=8.5cm]{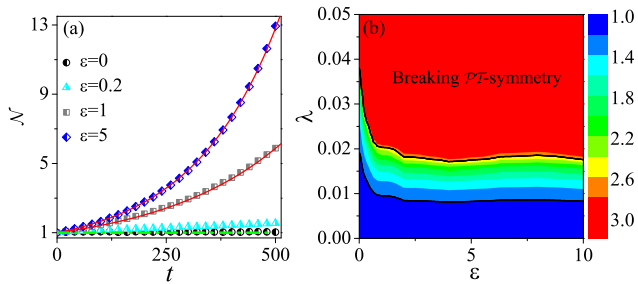}
\caption{(a) Norm $\mathcal{N}$ versus time with $\lambda=0.01$ for $\varepsilon=0$ (circles), 0.2 (triangles), 1 (squares), and 5 (diamonds).  Red solid lines indicate the fitting function of the form $\mathcal{N}\propto e^{\gamma t}$ with $\gamma = 0.0035\;(0.0051)$ for $\varepsilon=1$ (5). Green dashed line marks $\mathcal{N}=1$. (b) The time-averaged value of norm $\bar{\mathcal{N}}$ in the parameter space ($\lambda$, $\varepsilon$). The red (blue) area indicates the breaking (un-breaking) phase of $\mathcal{PT}$-symmetry. Other parameters are the same as in Fig.~\ref{Diffusion}.\label{Norm}}
\end{center}
\end{figure}

It is known that without interaction (i.e., $\varepsilon=0$) there is a threshold value for the imaginary part of the kicking potential, i.e., $\lambda_c$, beyond which the system is in the regime of the $\mathcal{PT}$-symmetry breaking phase. For convenience, the norm $\mathcal{N}(t_{n})=\sum_{m,n}|\psi_{m,n}(t_{n})|^2$~\cite{Zhou18a,Zhou18b,Longhi21} is applied to quantify the $\mathcal{PT}$-symmetry phase transition.
Figure~\ref{Norm}(a) shows that, for $\varepsilon=0$, the $\mathcal{N}$ remains at unity as time evolves, which demonstrates the maintenance of the $\mathcal{PT}$-symmetry phase. It is interesting that, for nonzero value of $\varepsilon$ (e.g., $\varepsilon=0.2$), the norm increases with time. Exponential growth of norm, i.e., $\mathcal{N}(t_{})\propto e^{\gamma t_{}}$, arises for sufficiently strong coupling (e.g., $\varepsilon=1$), which is a solid evidence of the spontaneous $\mathcal{PT}$-symmetry breaking. Therefore, the inter-particle coupling dramatically alters the phase transition point $\lambda_c$. To investigate the dependence of $\lambda_c$ on $\varepsilon$, we numerically calculate the time-averaged value of norm $\mathcal{\bar{N}}=(1/t_{M})\sum^M_{n=1}\mathcal{N}(t_{n})$ in the parameter space $(\lambda,\varepsilon)$. Figure~\ref{Norm}(b) shows that there are clearly two different regimes, corresponding to $\mathcal{\bar{N}}>1$ and $\mathcal{\bar{N}}=1$, respectively. Detailed observations show that the phase transition point $\lambda_c$ decreases with the increase of $\varepsilon$, which reveals the fact that the interaction is helpful to assist breaking the $\cal{PT}$-symmetry phase. By comparison with Fig.~\ref{PDiagram}, one can find that classes III and IV of quantum transport is in the region of the breaking phase of $\cal{PT}$-symmetry.

\subsection{Time evolution of the Linear entropy}

\begin{figure}[htbp]
\begin{center}
\includegraphics[width=8.5cm]{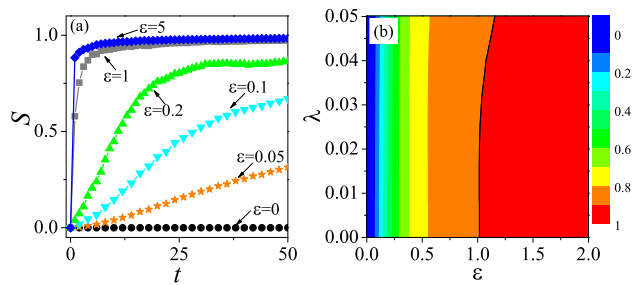}
\caption{(a) Linear entropy $S$ versus time for $\lambda=0.01$ with $\varepsilon=0$ (circles), 0.05 (pentagrams), 0.1 (down triangles), 0.2 (up triangles), 1 (squares), and 5 (diamonds). (b) Phase diagram of decoherence quantified by the time-averaged $\bar{S}$ in the parameter space ($\lambda,\varepsilon$). Other parameters are the same as in Fig.~\ref{Diffusion}.\label{Entropy}}
\end{center}
\end{figure}

In the decoherence theory, the unavoidable coupling between system and environment leads to the formation of entanglement. After tracing out the degree of freedoms of environment, the quantum coherence in the state of system is destroyed, resulting in a mixed state~\cite{Zurek03,Schlosshauer04}. To quantify entanglement, we numerically investigate the time evolution of the linear entropy $S=1-\rm{Tr}(\rho_{1}^{2})$~\cite{Santos00,Buscemi07,Zarate11}. Figure~\ref{Entropy}(a) shows that, for a specific $\varepsilon$ (e.g., $\varepsilon=0.2$), the $S$ increases linearly with time until saturates, which demonstrates the formation of entanglement. The saturation value of $S$ increases with $\varepsilon$ up to the maximum value $S_{max} \approx 1$, representing the growth of entanglement with coupling strength. Furthermore, we numerically calculate the time-averaged value of the linear entropy, i.e., $\bar{S}=(1/t_{M})\sum^M_{n=1}S(t_{n})$ for a wide regime of $\varepsilon$ and $\lambda$. Figure~\ref{Entropy}(b) shows that for a specific $\lambda$ the $\bar{S}$ increases from zero to almost unity with the increase of $\varepsilon$, which is a solid confirmation of the enhancement of entanglement by coupling.

\begin{figure}[htbp]
\begin{center}
\includegraphics[width=7.5cm]{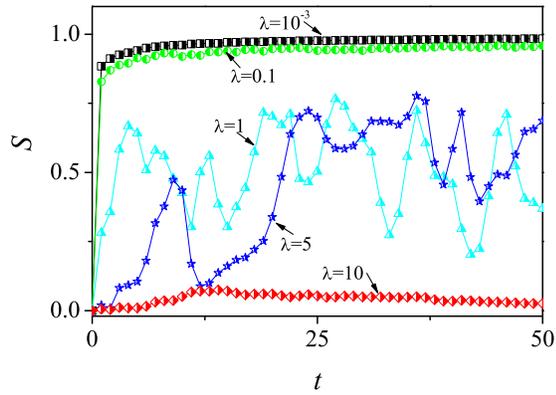}
\caption{Linear entropy $S$ versus time with $\varepsilon=5$ for $\lambda=10^{-3}$ (squares), $\lambda=0.1$ (circles), $\lambda=1$ (triangles), $\lambda=5$ (pentagrams), and $\lambda=10$ (diamonds). Other parameters are the same as in Fig.~\ref{Diffusion}.\label{Entropy2}}
\end{center}
\end{figure}

We further investigate the time evolution of the linear entropy $S$ for a wide regime of $\lambda$. Figure~\ref{Entropy2} shows that for small $\lambda$ (e.g., $\lambda\leq 0.1$) the $S$ increases very rapidly from zero to the saturation of almost unity, demonstrating the growth of entanglement in the coupled PTKRs. In this situation, the momentum distribution can be well described by the Gaussian function [see Fig.~\ref{Diffusion2}(d)], which is a character of the onset of the decoherence effects.
Interestingly, the saturation value of $S$ decreases with the increase of $\lambda$. For instance, the saturation of $S$ with $\lambda=5$ fluctuates around 0.5 as time evolves. The corresponding wavepacket differs clearly from the Gaussian function [see Fig.~\ref{Diffusion2}(d)], which may imply that the quantum coherence is partially protected by non-Hermitian driving.
For sufficiently large $\lambda$ (e.g., $\lambda=10$), the $S$ remains almost at zero with time evolution, that is a clear evidence of the disentanglement of the two particles due to the effects of the non-Hermitian driving~\cite{Huang21,KQHuang22}. It is reasonable to believe that the loss of quantum coherence is dramatically affected by the interplay between coupling and non-Hermitian driven potential.
This sheds a light on the issue of the quantum-classical transition induced by quantum decoherence effects in non-Hermitian chaotic systems.

\section{Conclusion and discussion}\label{SEC-SUM}

In this work, we study the transport behaviors in momentum space via a system of coupled PTKRs. Our investigations show that the phase transition point $\lambda_c$ of the spontaneous $\mathcal{PT}$-symmetry breaking can be modulated by the coupling strength $\varepsilon$. For $\lambda< \lambda_c$, the energy diffusion of each PTKR exhibits the transition from DL $\langle p^2\rangle \sim C$ [see phase I in Fig.~\ref{PDiagram}] to chaotic diffusion $\langle p^2\rangle \approx \gamma t$ [see phase II in Fig.~\ref{PDiagram}] with increasing the coupling strength. Meanwhile, there is no the emergence of the DC in momentum space. The DC $\langle p_1\rangle =Dt$ arises in the regime of $\cal{PT}$-symmetry breaking phase (i.e., $\lambda>\lambda_c$). At the same time, each PTKR exhibits the ballistic diffusion $\langle p^2_1\rangle\approx D^2t^2$ for weak coupling [see phase III in Fig.~\ref{PDiagram}], and the MBD $\langle p^2_1\rangle\approx D^2t^2 + \eta t^{\alpha}$ for strong coupling [see phase IV in Fig.~\ref{PDiagram}]. We numerically investigate the time evolution of linear entropy $S$, and find that for large $\varepsilon$ the $S$ increases rapidly from zero to saturation with time evolution. The saturation value of $S$ increases with increasing $\varepsilon$. We believe that the decoherence effects are responsible for the appearance of the intrinsic transport behaviors in the system of coupled PTKRs.

In recent years, the Floquet-driven systems~\cite{Shirley65,Bukova15} with periodical potential in time domain provide ideal platforms for investigating novel phenomena, such as quantum thermalization~\cite{Fleckenstein21,Geraedts16,Malishava}, many-body quantum chaos~\cite{Rylands20,Zhao14,Lundh06,Wang20}, and topologically-protected transport~\cite{Roy17,Fernandez19,Dag22}. The fate of DL and AL under the effects of interaction has received extensive investigations in the fields of quantum chaos and condensed matter physics. It is shown that temporally periodical-modulated nonlinearity even induces exponentially-fast diffusion in momentum space~\cite{WLZhao19,WLZhao20,Guarneri17}. Our finding of the coexistence of the DC and MBD in coupled PTKRs serves as an new element of quantum transport in non-Hermitian systems.

\textcolor{blue}{\textit{Acknowledgements}---}
Jian-Zheng Li is supported by the Dr. Start-up Fund of Jiangxi University of Science and Technology (No. 205200100067), and the Science and Technology Research Program of Jiangxi Education Department(No. GJJ190463). Wen-Lei Zhao is supported by the National Natural Science Foundation of China (Grant No.12065009) and Science and Technology Planning Project of Ganzhou City (Grant
No. 202101095077). Jie. Liu is supported by the NSAF (Contract No. U1930403).

\end{document}